# Quark Diagram Analysis of B-meson emitting vector (V) and vector (V) mesons

Maninder Kaur
*Department of Physics, Punjabi University,*
*Patiala – 147002, India.*

*e-mail:* maninderphy@gmail.com,

***Abstract***

This paper presents the two body weak nonleptonic decays of *B*-mesons emitting vector (*V*) and vector (*V*) mesons within the framework of the diagrammatic approaches at flavor *SU(3)* symmetry. We have investigated exclusive two body decays of *B*-meson using model independent quark diagram scheme. We have shown that the recent measurement of the two body exclusive decays of *B*-mesons can allow us to determine the magnitude and even sign of the QD amplitude for $B \rightarrow VV$ decays. Therefore, we become able to make few predictions for their branching fractions.

†Corresponding author: maninderphy@gmail.com

## 1. Introduction

On the theoretical side, various attempts had been made to investigate weak decays of heavy mesons. A number of technique has been employed. There is no rigorous method available at this moment for handling the necessarily non-perturbative physics of the exclusive *B* transitions. There exists another model independent so called 'quark diagram' or 'topological diagram' approach in the literature where all two-body non-leptonic weak decays of heavy mesons are expressed in terms of distinct quark diagrams, depending on the topologies of weak interactions, including all strong interaction effects. In future, a large quantity of new and accurate data on decays of the heavy flavor hadrons, is expected which calls for their theoretical analysis. Being heavy, bottom hadrons have revealed several channels for few decays, categorized as leptonic, semi-leptonic and hadronic decays [1-7]. Standard model provides satisfactory explanation of the leptonic and semileptonic decays but weak hadronic decays confronts serious problem as these decays experience strong interactions interferences. So our focus is to understand the weak hadronic decays of charm and bottom hadrons emitting *s*-wave.

Therefore, we investigate the two- body weak hadronic decays of heavy flavor hadrons emitting two vector mesons in the framework of standard model. This paper is the extension of our previous work where we had studied the $B \to PP$ [8] and $B \to PV$ decays [9]. For $B \to VV$ decays, several measurements on their branching ratios have been made, and more data is expected to become available in near future experiments. However, $B \to VV$ decays are more complicated than $B \to PP$ and $B \to PV$ because of the three possible partial waves in the final state. Experimental information at present is not detailed enough to separate the partial wave amplitudes. Most of the theoretical work has so far been devoted to understanding mainly their decay rate. The task is hampered for the computation of matrix elements between the initial and the final hadron states due to low energy strong interferences effects. In order to deal with these complicated matrix elements, usually the naïve and QCD factorization schemes [10-13] including probable final state interactions (FSI) [14-16] have been employed to predict branching fractions of nonleptonic decays of *B*-mesons. For some channels of *B* decays, however, the factorization calculations appear to be in clear disagreement with current measurement. The available data for two-body decays of heavy flavor mesons have indicated the presence of large nonfactorizable contributions [17-24], especially for the color suppressed decays. These decays have also been studied using flavor $SU(3)$ symmetry [25-28], where various dynamical factors get lumped into a few reduced matrix elements, which are generally determined using some experimental results. Fortunately, the experimental progress for weak semileptonic and nonleponic decays of the bottom mesons during the last years has been really astounding, due to which a good amount of experimental data now exists [7], which have inspired several theoretical works [1-5] on the weak decays of *B*-mesons.

In the present work, we have studied $B \to VV$ weak decays investigating contributions arising from various quark level weak interaction processes. Due to the strong interaction interference, like FSI and nonfactorizable contributions, on these processes, it is not possible to calculate their contributions reliably. For instance, weak annihilation and W-exchange contributions, which are naively expected to be suppressed in comparison to the W-emission terms, may become significant due to possible nonfactorizable effects arising through soft-gluon exchange around the weak vertex. Since such effects are not calculable from the first principles, we employ the model independent

Quark diagram approach, naively called Quark Diagram Scheme (hereafter referred to as QDS) [29-31]; wherein decay amplitudes (referred to as Quark amplitudes) can be expressed independently in terms of the topologies of possible quark flavor diagrams- like: a) the external W-emission diagram, b) the internal W-emission diagram, c) the W-exchange diagram, d) the W-annihilation, and e) the W-loop diagram, and parameterize their contributions to *B*-meson decays. The QDS has already been shown to be a useful technique for heavy flavor weak decays.

In section 2, we construct the weak Hamiltonian responsible for the $B \rightarrow VV$ decays. Choosing appropriate components of the weak quark level processes, we then obtain several straightforward relations among their decay amplitudes in Cabibbo-Kobayashi-Maskawa (CKM) enhanced as well as suppressed modes in section 3. In section 4, we proceed to derive corresponding relations among their branching fractions in the QDS using *SU(2)*-isospin, *SU(2)*-U spin, flavor *SU(3),* to relate many decay modes of heavy quarks. In this analysis, we have considered only those decays for which some experimental data exist to establish the applicability of the QDS. Consequently, predictions of some of the decay branching fractions are also made, which can provide further tests of the scheme. Summary and discussion are given in the last section.

## 2. Weak Hamiltonian

In the standard model, based on the gauge groups $SU(3)_C \otimes SU(2)_L \otimes U(1)_Y$, the quark couple to the W- boson through the weak current. The nonleptonic Hamiltonian has the usual current $\otimes$ current form

$$H_w = \frac{G_F}{\sqrt{2}} J_\mu^+ J^\mu + h.c, \tag{1}$$

where the weak current $J_\mu$ is given by

$$J_\mu \;=\; (\bar{u}\;\bar{c}\;\bar{t})\;\gamma_\mu(1\text{-}\gamma_5)\begin{pmatrix} d' \\ s' \\ b' \end{pmatrix}. \tag{2}$$

Weak eigenstates ($d'$, $s'$ and $b'$) are related to the mass eigenstates ($d$, $s$ and $b$) through the Cabibbo-Kobayashi-Maskawa mixing. We consider hadronic decay of *B*-mesons induced at the quark level by $b \rightarrow c/u$ transitions. The weak Hamiltonian generating the *b*-quark decays is thus given by

$$\begin{aligned} H_w^{\Delta b=1} = \frac{G_F}{\sqrt{2}} [\, & V_{ub} V_{cd}^* \,(\bar{u}\, b)(\bar{d}\, c) + V_{ub} V_{cs}^* \,(\bar{u}\, b)(\bar{s}c) + V_{ub} V_{ud}^* \,(\bar{u}b)(\bar{d}u) + \\ & V_{ub} V_{us}^* \,(\bar{u}b)(\bar{s}u) + V_{cb} V_{ud}^* \,(\bar{c}b)(\bar{d}u) + V_{cb} V_{us}^* \,(\bar{c}b)(\bar{s}u) + \\ & V_{cb} V_{cs}^* \,(\bar{c}b)(\bar{s}c) + V_{cb} V_{cd}^* \,(\bar{c}b)(\bar{d}c)\, ]. \end{aligned} \tag{3}$$

The color and space-time structure is omitted. Selection rules for various decay modes generated by the Hamiltonian are given below.

(i) CKM enhanced modes: $\Delta C = 1, \Delta S = 0; \Delta C = 0, \Delta S = -1;$

(ii) CKM Suppressed modes: ΔC = 1, ΔS = -1; ΔC = 0, ΔS = 0;
(iii) CKM doubly suppressed modes: ΔC = ΔS = -1; ΔC = -1, ΔS = 0.
Since only quark fields appear in the Hamiltonian, the *B*-meson decays are seriously affected by the strong interactions. One usually identifies the two scales in these decays: short distance scale at which W-exchange takes place, and long distance scale where final state hadrons are formed. The short distance effects are calculable using the perturbative QCD, which are expressed in terms of certain QCD coefficients. The long-distance effects being non-perturbative are the source of major problems in obtaining the decay amplitudes from the Hamiltonian, even after including the short distance modifications [32-33].

There are many ways that the quarks produced in a weak nonleptonic process can arrange themselves into final state hadrons. All *B*-meson decays can be expressed in terms of a few quark level diagrams [29-31]: a) the external W-emission diagram, b) the internal W-emission diagram, c) the W-exchange diagram, d) the W-annihilation, and e) the W-loop penguin diagram. Initially, it was expected that W-exchange and W-annihilation diagrams are suppressed due to the helicity and color considerations, and the penguin diagrams, involving W-loop, contribute to only two out of the six decay modes. Thus the dominant quark level process apparently seems to involve W-emission, in which light quark in the *B*-meson behaves like spectator. However, measurements of some of *B*-meson decays have challenged this naïve and simple picture, and it is now established that the non-spectator contributions may play a significant role in understanding the weak decays of heavy flavor hadrons. In fact, exchange of the soft gluons around the weak vertex also enhances such non-spectator contributions from the W-exchange, W-annihilation and W-loop diagrams. Unfortunately, these effects, being non-perturbative, cannot be determined unambiguously from the first principles.

In the absence of the exact dynamical calculations, we have employed the QDS to investigate contributions from different weak quark level diagrams. Such a scheme gives model independent way to analyze data to test the mechanism of the various quark level processes, and to make useful predictions for the meson decays. The decay amplitudes are obtained using the valence quark structure of the particles involved in the *B*-meson decays. Using the tensorial notation, the decay amplitudes are then obtained from the following contractions:

$$\begin{aligned} H_w^{\Delta b=1} &= [a(B^m V_m^i V_n^k) + d(B^i V_n^m V_m^k)]\, H_{[i,k]}^n \\ &+ [a'(B^m V_m^i V_n^k) + d'(B^i V_n^m V_m^k)]\, H_{(i,k)}^n \\ &+ [c(B^n V_n^m V_m^i)]\, H_i . \end{aligned} \tag{4}$$

The brackets [,] and (,) respectively, denote antisymmetrization and symmetrization among the indices *i, k.* In the flavor *SU(4)*, the *b*-quark behaves like singlet, and *u, d, s,* and *c* quarks form a quartet. Thus Hamiltonian for $\Delta b$ = 1 weak process belongs to the representations appearing in

$$4^* \otimes 1 \otimes 4^* \otimes 4 = 4^* \oplus 4^* \oplus 20' \oplus 36^*. \tag{5}$$

The weak spurion $H_i$, $H_{[i,k]}^n$ and $H_{(i,k)}^n$ belong to the 4*, 20', and 36* representations, respectively. However, we do not use the complete *SU(4)*-QDS due to *SU(4)* being badly broken, and exploit the QDS at the *SU(3)* level through the following *SU(4)*$\rightarrow$ *SU(3)* decomposition:

$$\begin{aligned} &4^* \supset 3^* \oplus 1, \\ &20' \supset 8 \oplus \left(6 \oplus 3^*\right) \oplus 3, \\ &36^* \supset 6^* \oplus \left(15 \oplus 3^*\right) \oplus \left(8 \oplus 1\right) \oplus 3. \end{aligned} \tag{6}$$

The *SU(3)*-QDS relates $\Delta C = 1$, $\Delta S = 0$ mode with $\Delta C = 1$, $\Delta S = -1$**;** $\Delta C = 0$, $\Delta S = -1$ mode with $\Delta C = 0$, $\Delta S = 0$; and $\Delta C = \Delta S = -1$ mode with $\Delta C = -1$, $\Delta S = 0$. The tensor $B^i$ denotes the parent *B*-mesons:

$$B^1 = B^+(\bar{b}u),\ B^2 = B^0(\bar{b}d),\ B^3 = B_s^0(\bar{b}s),\ B^4 = B_c^+(\bar{b}c). \tag{7}$$

The $V^i_j$ denote $4^* \otimes 4$ matrices of bottomless Vector mesons,

$$V^i_j = \begin{pmatrix} V^1_1 & \rho^+ & K^{*+} & \overline{D^{*0}} \\ \rho^- & V^2_2 & K^{*0} & D^{*-} \\ K^{*-} & \overline{K^{*0}} & V^3_3 & D^{*-}_s \\ D^{*0} & D^{*+} & D^{*+}_s & V^4_4 \end{pmatrix}. \tag{8}$$

Using *SU(3)* nonet (or *SU(4)* sixteenplet) symmetry, the diagonal states are taken to be:

$$V^1_1 = \frac{\rho^0 + \omega}{\sqrt{2}},\ V^2_2 = \frac{-\rho^0 + \omega}{\sqrt{2}},\ V^3_3 = -\phi,\ V^4_4 = \psi, \tag{9}$$

for ideal $\omega - \phi - \psi$ mixing $\vartheta_V = 0$. Quark content of the physical mesons [4, 7] are given by:

$$\rho^0 = \frac{1}{\sqrt{2}}(u\bar{u} - d\bar{d}),\ \omega = \frac{1}{\sqrt{2}}(u\bar{u} + d\bar{d}), \quad \phi = s\bar{s}, \quad \psi = c\bar{c}. \tag{10}$$

There exists a straight correspondence between the terms appearing in (4) and various quark level diagrams. The terms with coefficients *(a + a')* represent external W-emission, *(a - a')* represent internal W-emission, the terms with coefficients *(d - d')* represent W-exchange and *(d + d')* for W-annihilation processes. The last term having coefficient *c* represents the W-loop penguin diagram contributions. In addition, all the QCD effects have been absorbed in these parameters, the following contractions may also be constructed in the light of nonet (or sixteenplet) symmetry:

$$\begin{aligned} &+[h(B^i V^k_n V^m_m)]\, H^n_{[i,k]} + [h'(B^i V^k_n V^m_m)]\, H^n_{(i,k)} \\ &+[f(B^i V^m_n V^n_m) + f'(B^i V^m_m V^n_n) + f''(B^n V^i_n V^m_m)]\, H_i\,. \end{aligned} \tag{11}$$

However, these terms correspond to OZI violating diagrams which are expected to be suppressed, and hence are ignored in the present scheme.

## 3. Decay Amplitudes Relations

Choosing the relevant components of the Hamiltonian given in (3), we obtain the decay amplitudes in the *SU(3)* QDS for $B\rightarrow VV$ for which some experimental results are available. Depending upon the weak quark level processes involved in these decays, we have categorized their relations given below in three different ways:

1) Only single weak process (W-emission or W-exchange or W-annihilation) contributes, (*A*, *B*, *C* and *D*);
2) Combination of two weak processes (other than penguin diagram) contributes, (*E* and *F*);
3) Combination of penguin diagram with other weak processes contributes, (*G*).

### A. W-external emission:

$$A(B_s^0\rightarrow \rho^+ D^*{}_s^-) = V_{ud} / V_{us}\, A(B^0\rightarrow K^{*+} D^{*-}) \quad (12)$$

$$A(B_s^0\rightarrow K^{*-} D^{*+}) = V_{cd} / V_{cs}\, A(B^0\rightarrow \rho^- D^*{}_s^+) \quad (13)$$

$$A(B^+\rightarrow \rho^0 D^*{}_s^+) = (1/\sqrt{2})\, A(B^0\rightarrow \rho^- D^*{}_s^+) \quad (14)$$

$$A(B^+\rightarrow \rho^0 D^*{}_s^+) = A(B^+\rightarrow \omega\, D^*{}_s^+) \quad (15)$$

### B. W-internal emission:

$$A(B^+\rightarrow K^{*+}\psi) = A(B^0\rightarrow K^{*0}\psi) \quad (16)$$

$$A(B^+\rightarrow \rho^+\psi) = (-\sqrt{2})\, A(B^0\rightarrow \rho^0\psi) \quad (17)$$

$$A(B^0\rightarrow \omega\,\psi) = (1/\sqrt{2})(V_{cd} / V_{cs})\, A(B^+\rightarrow K^{*+}\,\psi) \quad (18)$$

$$A(B_s^0\rightarrow \bar{K}^{*0}\psi) = (-V_{cd} / V_{cs})\, A(B_s^0\rightarrow \varphi\,\psi) \quad (19)$$

$$A(B_s^0\rightarrow \bar{K}^{*0}\,\bar{D}^{*0}) = V_{ud} / V_{su}\, A(B^0\rightarrow K^{*0}\,\bar{D}^{*0}) \quad (20)$$

$$A(B_s^0\rightarrow \omega\,\varphi) = A(B_s^0\rightarrow \rho^0\,\varphi) \quad (21)$$

$$A(B_s^0\rightarrow \varphi\, D^{*0}) = A(B^0\rightarrow K^{*0} D^{*0}) \quad (22)$$

$$A(B_s^0\rightarrow \varphi\,\bar{D}^{*0}) = -A(B^0\rightarrow K^{*0}\,\bar{D}^{*0}) \quad (23)$$

$$A(B_s^0\rightarrow \bar{K}^{*0} D^{*0}) = V_{cd} / V_{cs}\, A(B^0\rightarrow K^{*0} D^{*0}) \quad (24)$$

### C. W-annihilation only:

$$A(B^+\rightarrow D^*{}_s^+\,\bar{K}^{*0}) = (-V_{cd} / V_{cs})\, A(B^+\rightarrow D^*{}_s^+\varphi) \quad (25)$$

$$A(B^+\rightarrow D^{*+} K^{*0}) = V_{cs} / V_{cd}\, A(B^+\rightarrow D^*{}_s^+\,\bar{K}^{*0}) \quad (26)$$

### D. W-exchange only:

$A(B^0 \to K^{*-}K^{*+}) = (\sqrt{2})(V_{ud}/V_{us})\, A(B_s^0 \to \rho^0\rho^0)$ (27)

$A(B_s^0 \to D_s^{*-}D_s^{*+}) = A(B^0 \to \bar{D}^{*0}D^{*0})$ (28)

$A(B_s^0 \to \bar{D}^{*0}D^{*0}) = V_{sc}/V_{dc}\, A(B^0 \to \bar{D}^{*0}D^{*0})$ (29)

$A(B_s^0 \to D^{*-}D^{*+}) = V_{sc}/V_{dc}\, A(B^0 \to \bar{D}^{*0}D^{*0})$ (30)

$A(B_s^0 \to \rho^0\omega) = (\sqrt{2})\, A(B_s^0 \to \rho^0\rho^0)$ (31)

$A(B_s^0 \to \omega\,\omega) = A(B_s^0 \to \rho^0\rho^0)$ (32)

$A(B_s^0 \to \rho^-\rho^+) = \sqrt{2}A(B_s^0 \to \rho^0\rho^0)$ (33)

**E. *W-emission (Both):***

$A(B^+ \to \bar{D}^{*0}K^{*+}) = V_{us}/V_{ud}\, A(B^+ \to \bar{D}^{*0}\rho^+)$ (34)

**F. *W-external emission and W- exchange:***

$A(B_s^0 \to K^{*+}D_s^{*-}) = V_{us}/V_{ud}\, A(B^0 \to D^{*-}\rho^+)$ (35)

**G. *W-emission (Both), W- annihilation and Penguin :***

$A(B^+ \to \rho^0K^{*+}) = A(B^+ \to \omega K^{*+})$ (36)

Note that the relations (14, 16, 17, 33) follow from the *SU(2)*- isospin framework, and (12-13, 20, 24, 26, 29, 30, 34, 35) from the *SU(2)*- U spin for QDS.

**4. Relations and Predictions for Branching Fractions:**

The decay rate formula for $B \to VV$ has the generic form:

$$\Gamma(B \to VV) = (non\text{-}kinematic\ factors)^2 \times \left(\frac{k}{8\pi M_B^2}\right) \times \left| \sum_{i=s,p,\mathrm{d}} decay\ amplitude\,(\mathrm{i}) \right|^2, \quad (37)$$

The following kinematic coefficients $\alpha_s$, $\alpha_P$, $\alpha_d$ and $\alpha_{s-d}$ :

$$\alpha_s = \left(2.0 + \frac{M_B^2 - m_1^2 - m_2^2}{2m_1m_2}\right)^2, \quad \alpha_P = \frac{8M_B^2k^2}{(\mathrm{M}_B + m_1)^4},$$

$$\alpha_d = \frac{8M_B^4k^4}{m_1m_2\,(M_B + m_1)^4}, \quad \alpha_{s-d} = \frac{2M_B^2k^2}{(M_B + m_1)^4} \cdot \frac{M_B^2 - m_1^2 - m_2^2}{2m_1m_2}. \quad (38)$$

control the relative contributions of the partial waves and the interference term. where $k$ is the 3-momentum of the final states and is given by

$$k = |p_1| = |p_2| = \frac{1}{2M_B}\left\{\left(M_B^2 - (m_1 + m_2)^2\right)\left(M_B^2 - (m_1 - m_2)^2\right)\right\}^{1/2} . \quad (39)$$

The hierarchy of the partial wave is expected to follow $|s| > |p| > |d|$ and the interference term is, consequently, small. The CLEO Collaboration [34] has performed the first full angular analysis of $B \to J/\psi\, K^{*0}$ and $J/\psi\, K^{*+}$ decays, which has confirmed that the *p*-wave component is significantly small in these decays [34-35]. It has been noted that the neglect of *p*- and *d*-partial waves would only reduce the decay rates by ~ (5–12) %. thus the retention of the *s*-wave only appears to be a reasonable approximation for $B \to VV$ decays mode. The decay rate is then given by

$$\Gamma(B \to VV) = \left(\frac{k}{8\pi M_B^2}\right)\left[2 + \left(\frac{M_B^2 - m_1^2 - m_2^2}{2m_1 m_2}\right)\right] |A(B \to VV)|^2 . \quad (40)$$

Several relations are obtained between branching fractions of the decays of $B^+$, $B^0$ and $B_s^0$ mesons, corresponding to the decay amplitude relations given in the previous section. We have used the available experimental values to check the consistency of the relations obtained and to predict the branching fractions of some of the decay not observed so far. We give our values just below the branching relations. These values are obtained by multiplying the known experimental value with the factor (given on RHS). For instance, in relation (41), branching fraction of $B(B_s^0 \to \rho^+ D^*{}_s^-)$ is obtained by multiplying the experimental value of branching fraction $B(B^0 \to K^{*+} D^{*-})$= (3.3 ± 0.6)× 10 $^{-4}$ with the factor 20.48 given in (41). Similar to the decay amplitude relations, we have categorized relations among the branching fractions according to the contributions arising from one or more of the weak quark diagrams. We have distinguished the $b \to s$ penguin process from that of $b \to d$.

### A. *W-external emission:*

$$B(B_s^0 \to \rho^+ D^*{}_s^-) = 20.48\, B(B^0 \to K^{*+} D^{*-}) \quad (41)$$
$(9.7 \pm 2.2)\times 10^{-3}$ $\quad$ $(6.8 \pm 1.2)\times 10^{-3}$

$$B(B_s^0 \to K^{*-} D^{*+}) = 0.05\, B(B^0 \to \rho^- D^*{}_s^+) \quad (42)$$

$$B(B^+ \to \rho^0 D^*{}_s^+) = 0.53\, B(B^0 \to \rho^- D^*{}_s^+) \quad (43)$$
$< 4\times 10^{-4}$ $\quad$ $(0.22 \pm 0.07)\times 10^{-4}$

$$B(B^+ \to \rho^0 D^*{}_s^+) = B(B^+ \to \omega\, D^*{}_s^+) \quad (44)$$
$< 4\times 10^{-4}$ $\quad$ $<6\times 10^{-4}$

### B. *W-internal emission:*

$$B(B^+ \to K^{*+} \psi) = 1.06\, B(B^0 \to K^{*0} \psi) \quad (45)$$
$(1.43 \pm 0.03)\times 10^{-3}$ $\quad$ $(1.39 \pm 0.06)\times 10^{-3}$

$$B(B^+ \to \rho^+ \psi) = 2.12\, B(B^0 \to \rho^0 \psi) \quad (46)$$
$(5.0 \pm 0.8)\times 10^{-5}$ $\quad$ $(5.38 \pm 0.29)\times 10^{-5}$

$$B(B^0 \to \omega\, \psi) = 0.03\, B(B^+ \to K^{*+}\psi) \quad (47)$$
$$(1.8^{+0.7}_{-0.5})\times 10^{-5} \qquad (4.29 \pm 0.09)\times 10^{-5}$$

$$B(B_s^0 \to \bar{K}^{*0}\psi) = 0.01\, B(B_s^0 \to \varphi\, \psi) \quad (48)$$
$$(1.08 \pm 0.09)\times 10^{-5}$$

$$B(B_s^0 \to \bar{K}^{*0}\, \bar{D}^{*0}) = 19.37\, B(B^0 \to K^{*0}\, \bar{D}^{*0}) \quad (49)$$
$$< 1.3\times 10^{-3}$$

$$B(B_s^0 \to \omega\, \varphi) = 0.99\, B(B_s^0 \to \rho^0\, \varphi) \quad (50)$$
$$< 6.05\times 10^{-4}$$

$$B(B_s^0 \to \varphi\, D^{*0}) = 7.58\, B(B^0 \to K^{*0}\, D^{*0}) \quad (51)$$
$$< 3.03\times 10^{-4}$$

$$B(B_s^0 \to \varphi\, \bar{D}^{*0}) = 7.58\, B(B^0 \to K^{*0}\, \bar{D}^{*0}) \quad (52)$$
$$< 5.2\times 10^{-4}$$

$$B(B_s^0 \to \bar{K}^{*0} D^{*0}) = 0.05\, B(B^0 \to K^{*0} D^{*0}) \quad (53)$$
$$< 0.21\times 10^{-5}$$

**C. *W-annihilation only:***

$$B(B^+ \to D_s^{*+}\, \bar{K}^{*0}) = 0.07\, B(B^+ \to D_s^{*+}\varphi) \quad (54)$$
$$<35\times 10^{-5} \qquad <0.08\times 10^{-5}$$

$$B(B^+ \to D^{*+} K^{*0}) = 20.58\, B(B^+ \to D_s^{*+}\, \bar{K}^{*0}) \quad (55)$$
$$<7.2\times 10^{-3}$$

**D. *W-exchange only:***

$$B(B^0 \to K^{*-} K^{*+}) = 0.16\, B(B_s^0 \to \rho^0 \rho^0) \quad (56)$$
$$<0.02\times 10^{-4} \qquad <0.51\times 10^{-4}$$

$$B(B_s^0 \to D_s^{*-}\, D_s^{*+}) = 0.09\, B(B^0 \to \bar{D}^{*0}\, D^{*0}) \quad (57)$$
$$<2.4\times 10^{-4} \qquad <0.08\times 10^{-4}$$

$$B(B_s^0 \to \bar{D}^{*0} D^{*0}) = 20\, B(B^0 \to \bar{D}^{*0} D^{*0}) \quad (58)$$
$$<1.8\times 10^{-3}$$

$$B(B_s^0 \to D^{*-} D^{*+}) = 20\, B(B^0 \to \bar{D}^{*0} D^{*0}) \quad (59)$$
$$<1.8\times 10^{-3}$$

$$B(B_s^0 \to \rho^0 \omega) = 3.96\, B(B_s^0 \to \rho^0 \rho^0) \quad (60)$$
$$<12.67\times 10^{-4}$$

$$B(B_s^0 \to \omega\, \omega) = 0.98\, B(B_s^0 \to \rho^0\, \rho^0) \quad (61)$$
$$<3.13\times 10^{-4}$$

$$B(B_s^0 \to \rho^- \rho^+) = 3.99\, B(B_s^0 \to \rho^0 \rho^0) \qquad (62)$$
$$< 12.77 \times 10^{-4}$$

**E. *W-emission (Both):***

$$B(B^+ \to \bar{D}^{*0} K^{*+}) = 0.05\, B(B^+ \to \bar{D}^{*0} \rho^+) \qquad (63)$$
$$(8.1 \pm 1.4) \times 10^{-4} \qquad (4.9 \pm 0.9) \times 10^{-4}$$

**F. *W-external emission and W- exchange: done***

$$B(B_s^0 \to K^{*+} D_s^{*-}) = 0.04\, B(B^0 \to D^{*-} \rho^+) \qquad (64)$$
$$(2.7 \pm 0.4) \times 10^{-4}$$

**G. *W-emission (Both), W- annihilation and Penguin:***

$$B(B^+ \to \omega K^{*+}) = B(B^+ \to \rho^0 K^{*+}) \qquad (65)$$
$$< 7.5 \times 10^{-6} \qquad (4.6 \pm 1.1) \times 10^{-6}$$

We note that almost all the relations obtained here are generally found to be consistent with experimental values. It may be remarked that the relations (41) to (62) among the decays, involving only one weak process, remain unaffected by any change of phase of the decay amplitudes, which may arise due to elastic FSI. However, other kind of relations, (63) to (65) where two or more weak processes contribute, may be affected by the relative phase factor. Further, it may also be noted that the relations (43, 45-46, 62) follow from the QDS at isospin level, and hence are more reliable. However, branching fraction relations (47, 63), show deviation from the experimental values. Note that available branching fraction for $B_s^0$ decays are scaled with $B\,(\bar{b} \to B_s^0)$ [7].

## 5. Summary and Conclusions

Thanks to the experimental efforts to deduce the branching fractions for $B \to VV$ decays. Theoretically, these decays are usually studied using the factorization scheme which expresses the decay amplitudes in terms of certain meson decay constants and meson-meson form factors. However, this scheme is unable to explain the experimental results, even after including hard QCD effects and possible phase differences. This may happen because of possible soft gluon exchange effects around the weak vertex, which may enhance the contributions of W-exchange, W-annihilation and W-loop processes. Since these effects are not extractable from the first principles, we have investigated the *B*-meson decays employing the framework of QDS. Firstly, we have obtained decay amplitude relations among $B \to VV$ decays using *SU(2)*-isospin, *SU(2)*-U spin, and *SU(3)* for the QDS. Afterwards, relations among their corresponding branching fractions have been derived in Sec. 4, giving experimental results wherever available.

we conclude that:

- The relations (41, 45-46) are consistent with the experimental data within the errors. So QDS seems to hold good for these hadronic weak decays of *B*-mesons. The relations (45-46), based upon *SU(2)*- isospin are found to be more reliable.

- Using the available branching fraction of observed decays, we also predict the branching fraction of several decays in (42-43, 48, 64-65). Explicitly $B(B^+ \rightarrow \rho^0 D^*{}_s{}^+) = (0.22 \pm 0.07)\times 10^{-4}$, $B(B_s^0 \rightarrow K^{*+} D_s^{*-}) = (2.7 \pm 0.4)\times 10^{-4}$, $B(B_s^0 \rightarrow \bar{K}^{*0}\psi) = (1.08 \pm 0.09)\times 10^{-5}$, $B(B_s^0 \rightarrow K^{*-} D^{*+}) = (2.04 \pm 0.64)\times 10^{-6}$, $B(B^+ \rightarrow \omega K^{*+}) = (4.6 \pm 1.1)\times 10^{-6}$ etc. Measurement of branching fractions of these decays would help to ascertain the validity of QDS.

- There are certain decay relations (44, 54, 56-57), where only upper limits are available, for both sides and few decay relations (49-53, 55, 58-62), are obtained where upper limits are available for one of the decays. These results may be tested in future experiments.

- Relation (47) shows slight deviation, due to *SU(3)* breaking which may produce 10-15% deviation.

- It may further be noted that the experimental branching fractions for the relations (63) have the same order, though different magnitudes. The (63) decay is related through *SU(2)*-U spin, and contain contribution from possible phase differences between two weak process involved in these decays.

**Acknowledgment:**

I am grateful for strong support and guidance provided to me by R. C. Verma, who helped me for useful discussions and reading the manuscript.